\documentclass{PoS}
\usepackage{graphicx}
\usepackage{caption}
\usepackage{subcaption}
\usepackage{amsmath,amsfonts,amssymb}
\usepackage{graphicx}
\usepackage{subcaption}
\let\OLDthebibliography\thebibliography
\renewcommand\thebibliography[1]{
\OLDthebibliography{#1}
\setlength{\parskip}{0pt}
\setlength{\itemsep}{0pt plus 0.3ex}
}
\usepackage{nicefrac}
\usepackage{graphicx,wrapfig,lipsum}
\usepackage{lineno}

\title{A better communicator is always a better scientist, or the reason why every PhD student should engage in science outreach}

\ShortTitle{A better communicator is always a better scientist}

\author{Catherine De Clercq\\
      Vrije Universiteit Brussel}
\author{Jérôme De Schauwers\\
      Université libre de Bruxelles -  Département Inforsciences}
\author{\speaker{Gwenhaël de Wasseige}\\
      IIHE - Vrije Universiteit Brussel\\
      E-mail: \email{gdewasse@vub.ac.be}}

\author{Jef van Laer\\
      Vrije Universiteit Brussel - Science Outreach Office}

\abstract{The ability to communicate with all audiences is a skill that is rapidly becoming a must-have for any future scientist. As more physicists engage in communicating science to non-expert audiences, research shows that this experience helps them to get a better understanding of their own research and the impact on society, improves the perception of science by lay audiences and can also become an area of personal growth as a citizen. A recent deployment of a PhD student to the Amundsen Scott South Pole Station, as part of the IceCube Collaboration, provided a ready opportunity to spark interest. We present results of the efforts made by the Université libre de Bruxelles (ULB), the Vrije Universiteit Brussel (VUB) and the Interuniversity Institute for High Energies, IIHE (ULB-VUB), to introduce Belgian students and citizens to science and the life of a scientist. The essential parts of this program will be identified to show why the contributions of a PhD student to the organization of these activities are beneficial to the development of new skills as a scientist, but also to broaden the audiences and the impact of the local university and/or the specific research outreach program.}

\FullConference{35th International Cosmic Ray Conference\\
		 10-20 July, 2017\\
		 Bexco, Busan, Korea}

\begin{document}
\section*{Introduction}
These proceedings present a set of activities that took place during Fall-Winter 2016-2017 in Belgium. These activities, jointly coordinated by the Université libre de Bruxelles (ULB) and the Vrije Universiteit Brussel (VUB), were organized in the framework of the trip to the South Pole of G. de Wasseige, PhD student in Brussels and member of the IceCube collaboration. 
Section~\ref{def} gives a general definition of science communication that motivates the project described in these proceedings. The activities are presented in Section~\ref{90degrees} and their visible impacts and consequences will be detailed in Section~\ref{impact}.

\section{Science communication: A definition}\label{def}
Science communication may be defined as the use of appropriate skills, media, activities, and
dialogue to produce one or more of the following personal responses to science (the vowel analogy)~\cite{scicomdef}:
\begin{itemize}
\item Awareness
\item Enjoyment
\item Interest
\item Opinion-forming
\item Understanding of science
\end{itemize}
While the term often refers to communication of science-related topic to non-experts, the audience can be anyone in the society: other scientists, communicators such as journalists,  science aficionados or a layman audience.

\subsection{Reasons to engage in science communication}
There are many reasons to engage in science communication, and every communicator has his/her own motivation. These reasons can be divided in three different categories: motivation for the public, the scientific community and for the own benefit of the communicator. We list a few of them here as examples.
\paragraph{For the public}
Discussing science-related topics with the general public has the direct effect of engaging more people in science. A public talk or a hands-on activity based on the actuality for example constitutes a tool for a layman audience to understand this actuality and/or make its own opinion about it.   
The activities broadening the knowledge of the participants allow to support and improve science education, by offering different perspectives of teaching.

\paragraph{For the communicator itself}
Engaging in science communication improves the own understanding of the communicator. Indeed, concepts or ideas that look simple when used on a daily basis may reveal unexpected complexity if discussed with a non-expert audience. The contextualization of his/her own research also allows a better understanding of the implications and increases the self confidence.
The visibility of the communicator among its community also benefits from its engagement in science communication. 
Finally, this engagement further helps to develop communication skills that will be useful to express ideas in any other professional situation.

\paragraph{For the scientific community}
It is often thought that communication is at the expense of quality research. Engaging in science communication demonstrates that PhD students or senior researchers can initiate and take part in outreach activities while developing scientific projects and leading research in the field. 
A larger engagement from the science community will also secure continuing research support by providing systematic feedback on the scientific progress and needs to the general public and politics.


\section{\textit{90$^{\circ}$ South}, where research joins outreach}\label{90degrees}

\textit{90$^{\circ}$ South} was a set of activities organized in the framework of the recent trip of G. de Wasseige, PhD student at the Interuniversity Institute for High Energies (ULB-VUB), to the South Pole. As a member of the IceCube collaboration, she has been selected to visit the IceCube neutrino telescope~\cite{jinst} buried in the South Pole ice. 
The IceCube collaboration, in which Belgium is the third main contributor, is one of the most successful of its generation. In 2013, IceCube has received the 'breakthrough of the year' award for the first detection of cosmic neutrinos~\cite{hese}. Since then, in parallel to the search of cosmic neutrinos that would allow to identify the first neutrino source in our Universe, the collaboration has found new ways to use the detector, leading to competitive measurements of neutrino oscillation parameters~\cite{theta} and setting strong constraints on sterile neutrinos among others~\cite{sterile}. IceCube has also successfully joined the worldwide effort aiming at a multimessenger observation of the Universe~\cite{amon, realtime, kevin}.

The trip of G. de Wasseige was the occasion to communicate about the great science achieved by IceCube in general and about the Brussels group in particular. Using the attractiveness of Antarctica and IceCube to trigger curiosity, we have developed several activities adapted for different target audiences. The aim of our activities was to develop awareness of science among Belgian citizens.

\subsection{Your experiment
at the South Pole?}

This experiment contest was dedicated to primary and high school students. Open to science and non-science oriented students, the challenge was to design an experiment answering the question "Belgium-South Pole: what is the difference?". 
An international jury selected the 3 experiments that have been carried out at the South Pole. The results were presented during a science fair together with all the participating experiments as illustrated in Fig.~\ref{exp}.

The goal of this activity was multiple: 
\begin{itemize}
    \item introducing Antarctica, its extreme conditions, its peculiar fauna, geography and climate and sensitizing the students to the existence of this continent dedicated to science.
    \item	giving the opportunity to the students to design their own experiment, taking into account the differences between the environment in Belgium and at the South Pole. This was the only constraint to guide the students and allow them to go in the direction of their choice. The experiments could therefore have been in physics, geography, biology, medical science or social science.
    \item 	teaching how to write a scientific report, using a template based on actual reports written by researchers.
    \item introducing experimental bias, reproducibility of an experiment and systematic errors using the results of the winning experiments that have been carried out at the South Pole
    \item 	engaging students in science communication by inviting them to present their work, setup and results during an exhibition in front of other participants, parents and students and professors of the ULB and VUB. 
\end{itemize}

The project not only developed awareness and knowledge in several scientific fields with the participants (aged 10 - 14), but it also provided them with valuable insights into the methods and 'nature of science'.

\subsection{Antarcticards:
a greetings exchange}

This trans-generational activity was dedicated to schools and families. Since G. de Wasseige was spending the end-of-the-year celebrations at the South Pole, we proposed to Belgian citizens to send their greetings to scientists spending December at the South Pole. We invited the participants to design their own original postcard or to use an existing one to send a bit of Belgium to the end of the World. This activity, focusing on scientists life, was offering a framework for passive science communication. The participants were indeed invited to question themselves:  "why can I send a postcard there? What are the people at the South Pole doing?  In which conditions?". In exchange, they received a postcard sent from the South Pole, carrying a unique post stamp as show in Fig.~\ref{cartes}.


\subsection{Closer look from the South Pole}
\begin{figure}
    \centering
    \begin{subfigure}[b]{0.45\textwidth}
         \includegraphics[width=0.95\textwidth]{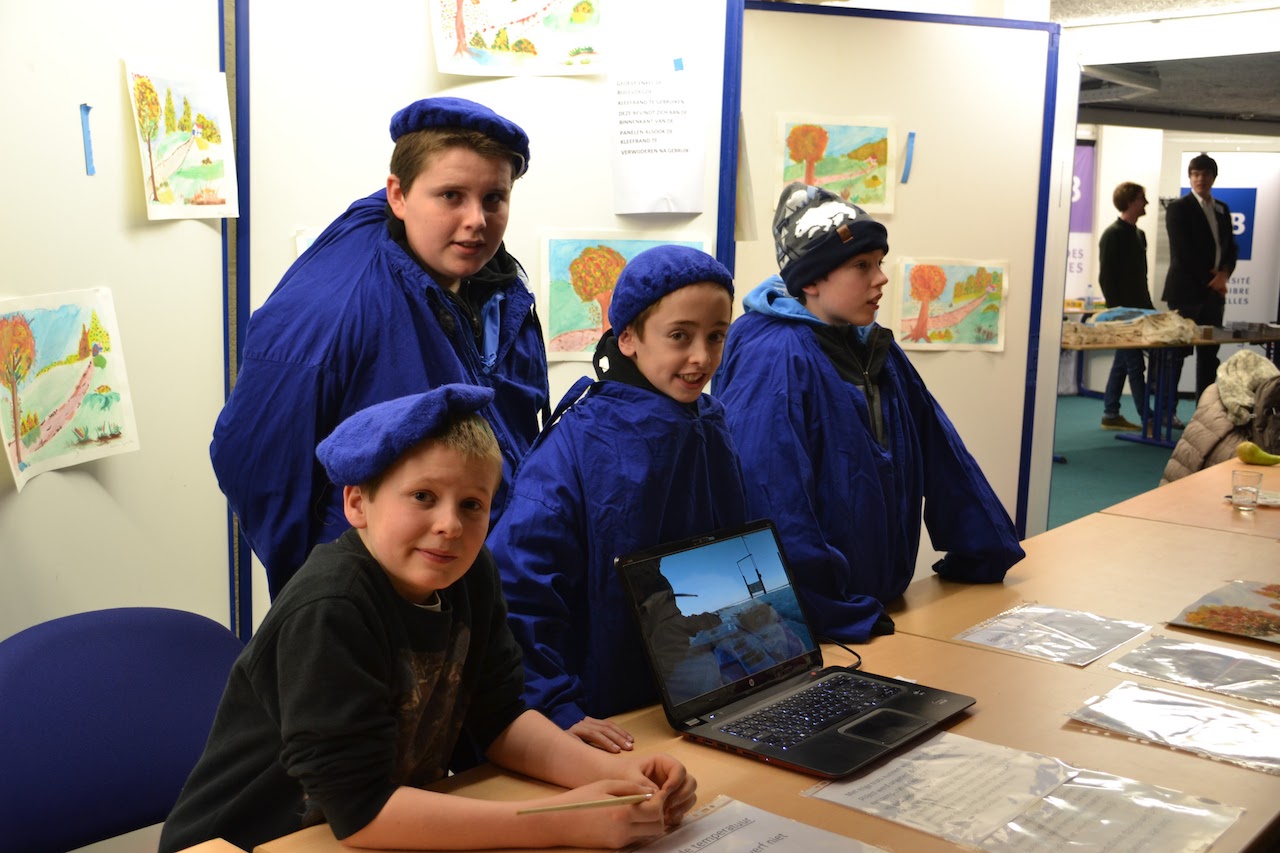}      
         \captionsetup{justification=centering}
         \caption{Students presenting their experiments during the science fair \scriptsize - G. de Wasseige IceCube/NSF}
        \label{exp}
    \end{subfigure}
   ~ \quad
    \begin{subfigure}[b]{0.45\textwidth}
         \includegraphics[width=0.95\textwidth]{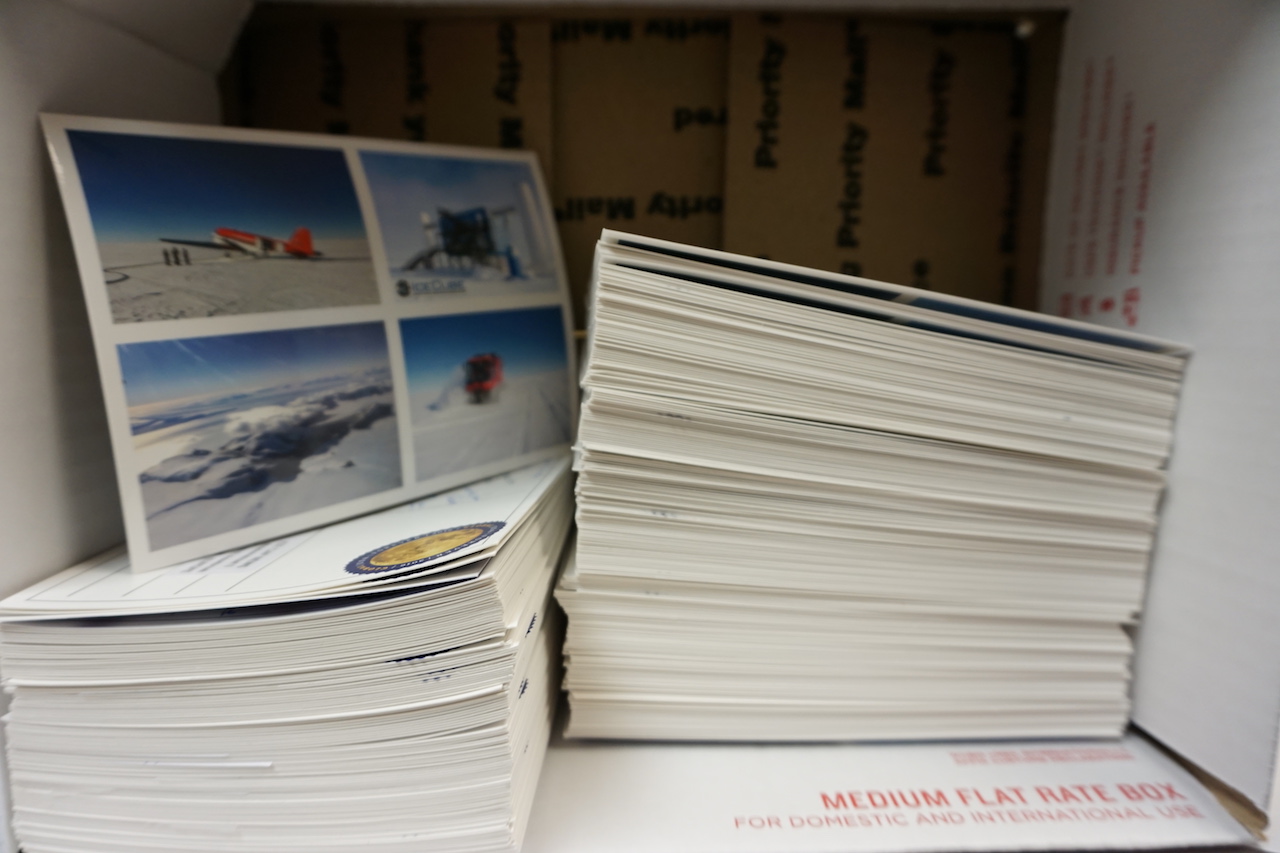}      \captionsetup{justification=centering}  \caption{Postcards sent from the South Pole to Belgian citizens \scriptsize - G. de Wasseige IceCube/NSF}
        \label{cartes}
    \end{subfigure}
    ~  \quad 
    
    \begin{subfigure}[b]{0.45\textwidth}
        \includegraphics[width=0.95\textwidth]{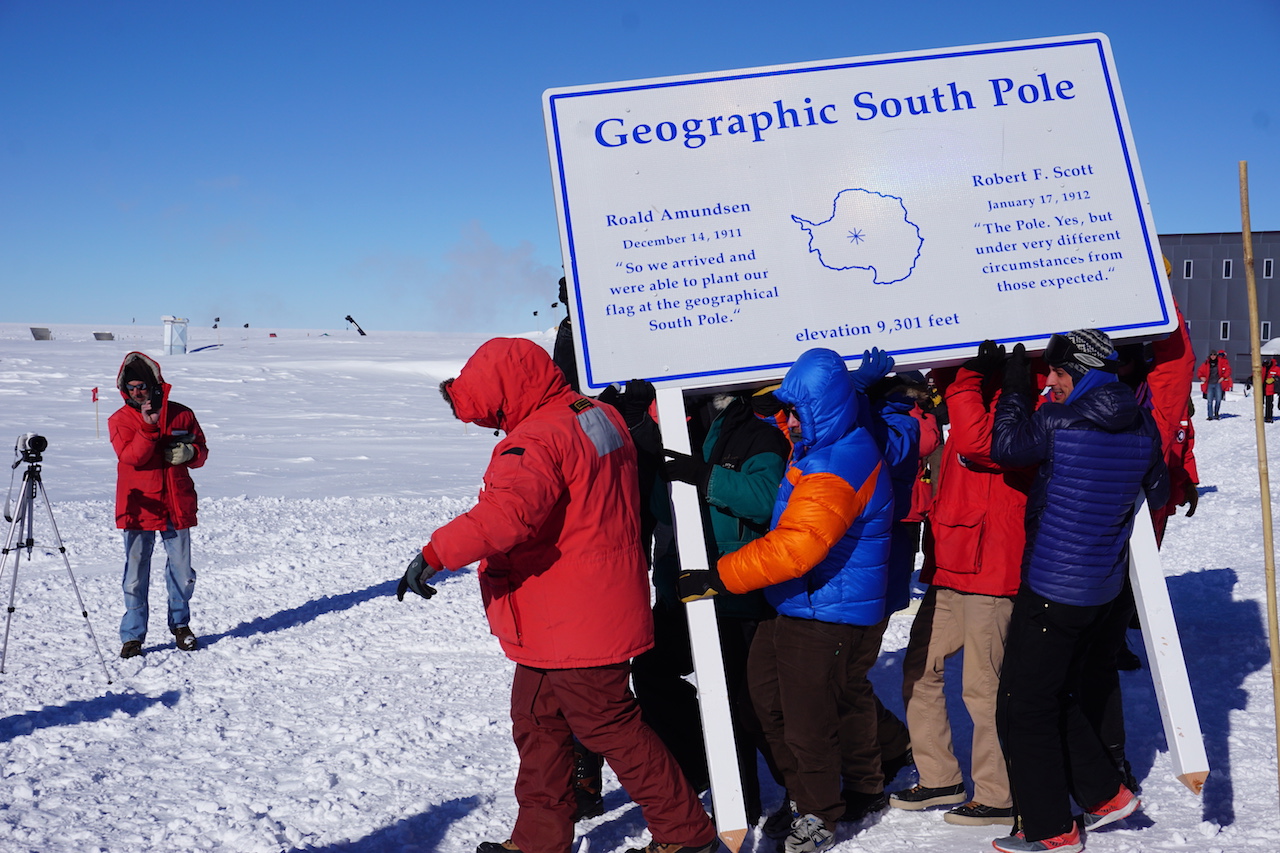}\captionsetup{justification=centering}
                \caption{Moving the South Pole \\ \scriptsize - G. de Wasseige IceCube/NSF}
        \label{twitter}
    \end{subfigure}
    \caption{}\label{fig:animals}
\end{figure}

Besides sharing her daily experience and impressions about the adventure on Twitter, G. de Wasseige organized a game challenging the social media community. She posted close up pictures of scientific material or South Pole related objects and asked the community to guess what it was and what it was used for.  The aim was to discuss, in a diverting way, the science carried out at the South Pole: atmospheric studies, astroparticle and astrophysics,  and highlight the global effort made to protect the environment from human effects.
The community following her was mainly composed of scientists or science-aficionados, curious about the work and experiences of other scientists. The pictures, e.g. Fig.~\ref{twitter}, used in the game were then printed on large scale canvas to form an exhibition targeting the same audience in Brussels.

\section{Impact of \textit{90$^{\circ}$ South}}\label{impact}
\subsection{Impact on society}
This project aimed to sensitize the public to science and especially science conducted by Belgian scientists. An additional goal was to interest the public to scientist's life and experience.
Through the experiment contest, we wanted to show that everyone can develop a scientific reasoning, even if not predisposed to pursue a scientific career.
We also wanted to prove that science can be brought outside of the classroom by presenting concrete situations to learn more about science. The collaborative and international aspects of science have also been illustrated through this project.
Finally, it is interesting to note that \textit{90$^{\circ}$ South} stands as an active learning project rather than as a talk-like program.

The expected number of participants for each activity has been reached or exceeded. The feedback received for each activity was better than one could have hoped. 
\begin{itemize}
\item \textit{Your experiment at the South Pole?:} More than 200 participants within 15 classrooms responded to the call. The students participating to the experiment contest were very enthusiastic and have shown a growing interest for science.
Several teachers ask whether we could  organize the contest again next year. All pointed out that the contest allowed to illustrate several important points of the school program by means of a concrete application.

It is also interesting to note that the exhibition closing the experiment contest was visited by numerous master and PhD students as well as professors, interested to be part of the jury committee evaluating the work of the participants. 
\item \textit{Antarcticards:}  More than 600 postcards have been received. Almost all of the postcards were illustrating/talking about the life at the South Pole. The participants were sensitive to the life conditions and the reasons why scientists were spending end-of-the-year celebrations in Antarctica.  Based on the content of the postcards, one can see that the adult participants realized the science carried out at the South Pole  while the young participants were more interested in the infrastructure in Antarctica as well as the fauna.
\item \textit{Closer look from the South Pole:}  The posts were on average seen by 1400 people, with a maximum of 4330\footnote{according to the statistics tool in Twitter}. 
\end{itemize}

In general the public was very enthusiastic about the adventure and the proposed activities. Several participants clearly expressed their wish to go to the South Pole and many have shown an increased interest for science in general and astrophysics in particular.
The local and national press extensively covered the different activities and interviewed G. de Wasseige before, during and after the trip.

\subsection{Personal impact}

As a PhD student and co-organizer of the \textit{90$^{\circ}$ South} project, G. de Wasseige has learned how to write a proposal to get funds and communicate with journalists, including live interviews and releases.  She also developed a presence on social media and improved science communication skills that she will be able to use in the future. 
This project in general raised awareness of IceCube and Belgian members of the collaboration among citizens, scientists and funding agencies.

\section{Summary}
These proceedings report on a set of activities that have been organized during Winter 2016-2017 In Belgium. These activities were elaborated around the trip to the South Pole of a PhD students, member of the IceCube collaboration.
Different audiences were the target of these activities, and they all show a positive response to the aim of this project: sensitization of science and scientist's life through the work of Belgian physicists. We also presented the impact on society as well as the personal impact one can expect when engaging in science communication.

As a conclusion, we would advise every scientist, student or senior researcher, to ask around for the opportunities, services and people at their own research institution that could support ideas. Research organisms, like universities or institutes, have resources and services available to support communication and outreach efforts, but it does require a researcher to come up with the idea and take the initiative.


\begin{thebibliography}{99}
\bibitem{scicomdef}
T.W. Burns, D.J. O'Connor and S.M. Stocklmayer, 
Public Understanding of Science {\bf12} (2003) 2, 183.
\bibitem{effectivescicom}
S. Illingworth and G. Allen, \textit{Effective Science Communication: A practical guide to surviving as a scientist}, IOP expanding physics, 2053-2563.
\bibitem{sciencecommunicator}
J. G. Radzilowicz, \textit{So,
You Want to be a Science Communicator?}, seminar.
\bibitem{icecube} 
IceCube communication workshops

\bibitem{jinst}
  M.~G.~Aartsen {\it et al.} 
  JINST {\bf 12} (2017) 03,  P03012.

\bibitem{deepcore}
R. Abbasi \textit{et al.}, 
 Astropart. Phys. \textbf{35} (2012) 615.

\bibitem{hese}
  M.~G.~Aartsen {\it et al.} 
  Science {\bf 342} (2013) 1242856.

 \bibitem{sterile}
  M.~G.~Aartsen {\it et al.}, 
  Phys.\ Rev.\ Lett.\  {\bf 117} (2016) 7,  071801.

\bibitem{theta}
 M.~G.~Aartsen {\it et al.},
  Nucl.\ Phys.\ B {\bf 908} (2016) 161.

\bibitem{amon}
  M.~W.~E.~Smith {\it et al.},
  Astropart.\ Phys.\  {\bf 45} (2013) 56.
\bibitem{realtime}
  M.~G.~Aartsen {\it et al.}, 
  Astropart.\ Phys.\  {\bf 92} (2017) 30.
\bibitem{kevin}
IceCube Coll., \pos{PoS(ICRC2017)1007} these proceedings.


\end{thebibliography}
\end{document}